\documentclass[conference]{IEEEtran}
\IEEEoverridecommandlockouts
\usepackage{balance}
\usepackage{array}
\usepackage{cite}
\usepackage{amsmath,amssymb,amsfonts}
\usepackage{algorithmic}
\usepackage{algorithm}
\usepackage{graphicx}
\usepackage{textcomp}
\usepackage{xcolor}
\usepackage{subcaption}
\usepackage{epstopdf}
\usepackage{enumerate}
\usepackage{multirow}

\def\BibTeX{{\rm B\kern-.05em{\sc i\kern-.025em b}\kern-.08em
    T\kern-.1667em\lower.7ex\hbox{E}\kern-.125emX}}
\begin{document}

\title{Double-Layer Game Based Wireless Charging Scheduling for Electric Vehicles}
\author{
	\IEEEauthorblockN{Tian Wang\IEEEauthorrefmark{1}\IEEEauthorrefmark{2}, Bo Yang\IEEEauthorrefmark{1}\IEEEauthorrefmark{2}, Cailian Chen\IEEEauthorrefmark{1}\IEEEauthorrefmark{2}}   
	\IEEEauthorblockA{\IEEEauthorrefmark{1}Department of Automation, Shanghai Jiao Tong University, Shanghai, China}  
	\IEEEauthorblockA{\IEEEauthorrefmark{2}Key Laboratory of System Control and Information Processing, Ministry of Education of China, China}
	\IEEEauthorblockA{Email: bo.yang@sjtu.edu.cn} 
	
}

\maketitle

\begin{abstract}

Wireless charging technology provides a solution to the insufficient battery life of electric vehicles (EVs). However, the conflict of interests between wireless charging lanes (WCLs) and EVs is difficult to resolve. In the day-ahead electricity market, considering the revenue of WCLs caused by the deviation between actual electricity sales and pre-purchased electricity, as well as endurance and traveling experience of EVs, this paper proposes a charging scheduling algorithm based on a double-layer game model. In lower layer, the potential game is used to model the multi-vehicle game of vehicle charging planning. A shortest path algorithm based on the three-way greedy strategy is designed to solve in dynamic charging sequence problem, and the improved particle swarm optimization algorithm are used to solve the variable ordered potential game. In the upper layer, the reverse Stackelberg game is adopted to harmonize the cost of wireless charging lanes and electric vehicles. As the leader, WCLs stimulate EVs to carry out reasonable charing action by electricity price regulation. As the follower, EVs make the best charging decisions for a given electricity price. An iteration algorithm is designed to ensure the Nash equilibrium convergence of this game. The simulation results show that the double-layer game model proposed in this paper can effectively suppress the deviation between the actual electricity sales and the pre-sale of the charging lane caused by the disorderly charging behavior of the vehicle, and ensure the high endurance and traveling experience of EVs. 
\end{abstract}


\section{Introduction}
\label{sct1}
With the increasing popularity of electric vehicles (EVs), the traditional plug-in charging station is difficult to meet the huge charging demand. In recent years, people have studied the application of wireless charging technology in EVs, which is expected to solve this problem. With this technology, EVs can charge on the wireless charging lane (WCL) while they are moving. However, due to the close relationship between EV dynamic charging behavior and traffic conditions, the irregular charging behavior of a large number of EVs is likely to have an adverse impact on power grid, such as power failure, excessive voltage fluctuation \cite{clement2010TheImpact}.
%To cope with the demand of electric vehicles, one solution is to increase power generation, which will lead to huge infrastructure costs. 
Therefore, the grid allows EVs to schedule their charging plan to improve the load condition of the grid.

There may be two energy flows between EVs and power grid, namely grid to vehicle (G2V) and vehicle to grid (V2G). V2G means EVs can charge or recharge on charging facilities. In recent years, many works have studied the V2G of plug-in EVs. In \cite{Abdulaal2017Solving}, G2V and V2G are combined in the path planning problem, and the hidden Markov model is used to model the charging behavior. Based on the improved genetic algorithm and multi-agent communication structure, the optimal path of EV is realized. In\cite{Liu2019Electric}, the research studies the reverse auction mechanism based on dynamic pricing strategy to complete transaction matching between buyer and seller EVs, improving the profit of sellers with weak competitiveness, and reduce the power purchase cost of buyers. 

According to whether the mobility of EVs is considered, the charging scheduling can be divided into immovable schedule and mobile schedule. In \cite{Saber2012Resource}, the optimization algorithm considering the uncertainty of renewable energy and load  is studied to reduce the cost and emission of power system. EVs are only regarded as uncertain loads of power grid in this work. The study \cite{Shafie2016Optimal} considers the temporal and spatial changes of buses and the accumulation of power demand in the electric public transportation system, and designs a time slot based electricity price scheduling strategy in the day ahead electricity market. It can be found that the research of immovable schedule mainly focuses on the impact of EV charging demand on the power grid, while mobile schedule can better reflect the traffic characteristics of vehicles.

In the wireless charging scenario, the WCLs are widely distributed, and vehicle charging is frequent, which cause load unbalance to the power grid. Hence V2G is not suitable. The charging condition of EVs varies with its route, and the dynamic charging sequence of WCLs makes the mobility of EVs negligible. The wireless charging plan of EVs will not only impact each other, but also affect the revenue of WCLs. This is the major difficulty of EV charging scheduling. In this paper, we design a double layer game model, EV-EV charging game as the lower game and WCLs-EVs game as the upper game, and corresponding algorithms to solve these problems. The contribution of this paper is as follows:
\begin{itemize}
	\item We convert the load balance requirement of power grid into WCL revenue by establishing the electricity price function based on the pre-purchased electricity in a day ahead electricity market, and build the congestion perception, charging value model, as well as the additional loss model of EVs;
	\item We use potential game to show the game balance existence in EV-EV charging game by designing a potential game. In addition, we design a Three way Greedy Shortest Path (TGSP) algorithm to calculate the dynamic charging sequence, as well as an improved Particle Swarm Optimization (PSO) algorithm to solve the potential game and get the best EV charging schedule.
	\item We use reverse Stackelberg game to model the WCLs-EVs game by sharing the electricity price function and design an iteration algorithm to obtain the optimal price in game equilibrium. 
\end{itemize}

The remainder of this paper is organized as follows. Section \ref{sct2} analyzes the relation of entities and related models of EVs and WCLs. Section \ref{sct3} shows double layer game model consisting of EV-EV game and EVs-WCLs game. The algorithm for resolving double layer game is provided in Section \ref{sct4}. Simulations to validate our results are given in Section \ref{sct5}. And Section \ref{sct6} provides the conclusion.

\section{Preliminaries And Problem Formulation}
\label{sct2}
\subsection{The Relation of Entities}
In order to better explain the problems in this work, we describe and analyze the interaction modes and mutual influences between entities in the vehicle wireless charging scenario.
\begin{itemize}
	\item \textit{Power Grid and WCLs:} 
	The energies of WCLs are purchased from the power grid. In a day ahead electricity market, WCLs submit the predicted power consumption of the next day to the power grid according to the historical information, and the power grid sets the power grid price accordingly. Because the wireless charging is closely related to the route of vehicles, WCLs are more likely to generate demand fluctuation, which will cause load instability and extra cost of the grid. In our problem, this is due to the inaccurate prediction of the pre-purchased power by WCLs, and the grid will transfer this loss to the purchase cost.
	\item \textit{WCLs and EVs:} As the energy supplier and service buyer, WCLs and EVs are the major entities of our work, whose strategies constitute the charging scheduling. We denote $I$ and $J$ as the set of EVs and WCLs respectively. The strategy of EVs have two aspects: whether to pass through a WCL $X_i = \{x^i_j, x^i_j \in \{0,1\}, \forall j \in J\}$ and the amount of charge in WCLs $U^i = \{u^i_j, u^i_j \in \mathcal{R}, \forall j \in J\}$, which directly affects the revenue and the additional loss of WCLs. By changing the price, the WCL directly affects its own electricity sales revenue and the charging cost of the electric vehicle, which is an effective way to stimulate the EVs to change their charging strategy.
	\item \textit{EV and EV:} In the transportation system, vehicles interact with each other, which is more prominent in the wireless charging scenario. First, the choice of WCLs will affect the traffic flow, and correspondingly affect the driving experience of other EVs. Moreover, the average power supply capacity of a WCL is affected by traffic flow \cite{Wireless2019Wang}, while the maximum charge quantity of an EV in the WCL is limited by this WCL's average power supply capacity, so the selection of WCLs indirectly affects the upper bound of the charge quantity of other vehicles. In addition, the EV charging affect the electricity sales of WCLs, which influences charging price, thus indirectly affecting the charging cost of other EVs in the same WCL.
%	\item \textit{Traffic Center and WCLs and EVs:}Because the wireless charging behavior of EVs has a great impact on traffic, it needs to be supervised by the traffic center. On the one hand, the center receives start and destination of each EV, and provides them with charging strategy recommendations. On the other hand,it obtains the predicted traffic flow and average vehicle speed according to the historical data and helps WCLs to plan pre-purchased electricity. Due to its rich traffic information collection and communication ability, the traffic center is the coordinator of WCLs and EVs.
\end{itemize}

\subsection{WCL Related Models}
\subsubsection{WCL Price Function}
\label{sec:price-function}
WCLs adjust the hourly electricity price to drive EVs to make charging plan that meets the supplies of WCLs. In the pricing strategy, it is necessary to consider the deviation between the pre-purchased electricity and the actual sold electricity to prevent the additional loss. Thus the price function consists of two parts: basic price and deviation price, as shown in \eqref{eq:price}.
\begin{equation}
\label{eq:price}
P(U_j)=p_0+q_j(U_j-\tilde{U_j}),
\end{equation}
where $p_0$ and $q_j$ are the basic price and the price coefficient, respectively. $U_j$ and $\tilde{U_j}$ denote the actual sold electricity and the predicted quantity, respectively.
\subsubsection{WCL Maximum Charge Quantity}
The maximum charge quantity of an EV is limited by real-time traffic condition. In our problem, we offer charging plans for multiple EVs without gathering real-time information of a single EV. So we define the average hourly charge amount of an EV as the WCL maximum charge quantity. In \cite{Wireless2019Wang} we can get the average hourly charge amount of WCL as \eqref{eq:maximal-charging-capacity}:
\begin{equation}
s = \dfrac{L_jf_j}{v_j},
\end{equation}
\begin{equation}
\xi = \dfrac{2\bar{d_2}}{v_{min}}-\dfrac{1}{f_j},
\end{equation}
\begin{equation}
\label{eq:maximal-charging-capacity}
U^{avr}_j = \dfrac{pL_jf_j}{v_j}+pn_js(T^r_j+\dfrac{(1+s)}{2}\xi),
\end{equation}
where $f_j$ and $v_j$ denote the traffic flow (vehicles per hour) and average vehicle speed of WCL $j$, respectively. $L_j$ and $p$ is the length and power of WCL $j$, respectively. $T^r_j$ denotes the average red traffic light duration at the WCL $j$ allocated intersection and $n_j$ denotes the number of traffic light cycles in an hour. $v_{min}$ denotes the minimum speed for leaving the intersection and $\bar{d_2}$ denotes the vehicle to vehicle distance between stopped EVs. $s$ and $\xi$ are intermediate variables. $U^{avr}_j$ denotes the average total charge of EVs in one hour on WCL $j$. 
%We can see that the relation of $U^{hour}_j$ and $f_j$,$v_j$ is complicated, while we do not schedule speed for EVs. 
For the convenience of analysis, we make an assumption to simplify the model:\textit{When the actual traffic flow is within the small range $\pm\omega$ of the predicted value, the average vehicle speed can be regarded as the same as the predicted value.} This assumption is meaningful for the reason that the pre-purchased electricity is based on the predicted traffic condition. If the deviation of actual and predicted traffic flow is too large, the actual sold electricity will deviate from the pre-purchased quantity. In addition, in order to maintain stable traffic flow, the traffic center will not allow large fluctuations in traffic flow. Hence we can simplify \eqref{eq:maximal-charging-capacity} into \eqref{eq:maximal-charging-capacity-2}:
\begin{equation}
\label{eq:maximal-charging-capacity-2}
U^{avr}_j = a_j f^2_j + d_j f_j + c_j,
\end{equation}
where $a_j$,$d_j$ and $c_j$ are constant coefficient. $c_j = -\dfrac{pn_j L_j}{2v_j}$, which means half of the charge obtained when the vehicle passes the WCL at speed $v_j$ multiplied by $n_j$, is a very small part that can be abandoned. Finally, we can conclude that the maximum charge quantity an EV can obtain from a WCL is linear to the traffic flow, as shown in \eqref{eq:single-charging-capacity}:
\begin{equation}
\label{eq:single-charging-capacity}
u_j(f_j) = \dfrac{U^{avr}_j}{f_j} = a_j f_j + d_j.
\end{equation}
\subsection{EV Related Models}
\label{subsec:ev-model}
\subsubsection{Additional Loss by Charging Route}
Different charging routes may generate additional traveling time and energy consumption, which is compared with the shortest route between the start and the destination of a vehicle. For the estimated time and energy consumption of the selected route, we multiply the route length by the corresponding coefficient, and convert it into economic loss, as \eqref{eq:additional-cost} and \eqref{eq:additional-time}:
\begin{equation}
\label{eq:additional-cost}
C(s,X,d) = p_0 \bar{p_c} (L(s,X,d)-L(s,d)),
\end{equation}
\begin{equation}
\label{eq:additional-time}
T(s,X,d) = p_1 \bar{p_t} (L(s,X,d)-L(s,d)),
\end{equation}
where $p_0$ and $p_1$ are the basic electricity price and the per capita hourly wage, respectively. $\bar{p_c}$ and $\bar{p_t}$ are the average energy and time cost per kilometer. $s$ and $d$ denote the start and destination respectively, and $X$ denotes the WCL choice. $L(s,d)$ calculates the shortest route length from the start to the end, and $L(s,X,d)$ calculates the length of the shortest route through the selected WCLs. We will talk about it in \ref{subsec:tgsp}.
\subsubsection{Congestion Perception}
The driving experience of vehicles is affected by road congestion. Such a congestion is closely related traffic flow, which is the major way for vehicles to percept congestion. The larger traffic flow is, the higher the degree of congestion that vehicles feel, and this feeling is stronger with more vehicles \cite{Etesami2017Smart}. Therefore, this congestion perception of vehicles is modeled as \eqref{eq:congestion-perception}:
\begin{equation}
\label{eq:congestion-perception}
D(f_j) = \varepsilon f^2_j,
\end{equation}
where $\varepsilon$ is the congestion perception coefficient. 
\subsubsection{Charging Value}
In the process of EV traveling, the driver hopes to charge as much as possible when passing through the WCLs, so as to increase the vehicle's endurance. Therefore, more charging quantity can bring greater benefits to an EV. We define this benefits as charging value and the charging value increases linearly with the charge quantity, as shown in \eqref{eq:charging-value}:
\begin{equation}
\label{eq:charging-value}
R(u^i_j) = \bar{p_r} u^i_j,
\end{equation}
where $\bar{p_r}$ is the converting coefficient and $u^i_j$ is the charge quantity of EV $i$ at WCL $j$. For different EV, the lower the initial state of charge (SOC) is, the greater $\bar{p_r}$ is.
\section{Double-Layer Game}
\label{sct3}
As mentioned above, WCLs' electricity price adjustment and EVs' charging decision will have mutual influence. WCLs want to improve the profit of electricity sales and EVs want to reduce the charging cost, which is the conflict of interest between them. For EVs, the conflicts between their decisions have also been discussed above. In order to solve the above two kinds of contradiction, we build a double-layer game model. The upper layer is the game between WCLs and EVs, and the lower layer is the game between EVs.

\subsection{Lower Layer: Game Between EVs}
As for the lower layer game, we take potential game to prove the existence of Nash equilibrium in the wireless charging game of EVs. Potential game maps the changes of individual cost to the common cost function by constructing the potential function. This model also simplifies the process in the upper layer game.

\subsubsection{EV Cost Function}
Every vehicle is a selfish individual, hoping to reduce the cost of purchasing electricity while ensuring sufficient power to travel, and avoiding detour. Therefore, we take the sum of EV losses in subsection \ref{subsec:ev-model} as the cost function of each EV, and get the objective function and constraints of an individual vehicle $i$ as follows:

\begin{align}
\min_{X_i,U^i} \sum_{j \in \left\{j | x_{j}^{i}=1\right\}}&\left[P\left(U^F_{j}\right) u_{j}^{i}+D\left(f_{j}\right)-R\left(u_{j}^{i}\right)\right] \nonumber\\
&+ C\left(s_{i}, X_{i}, d_{i}\right)+ T\left(s_{i}, X_{i}, d_{i}\right)                      
\label{obj:single-car}\\
{s.t.}\qquad\quad\  U^F_{j}&=\sum_{i \in I} u_{j}^{i},			\label{constraint:single-car-1}\\
f_{j}&=\sum_{i \in I} x_{j}^{i},														\label{constraint:single-car-2}\\
0 &\leq u_{j}^{i} \leq u_{j}(f_j),														\label{constraint:single-car-3}\\
u_{s_{k}}^{i} &\leq b_{m a x}-b^i_{s_{k}}, \quad s_{k} \in\left\{j | x_{j}^{i}=1\right\},	\label{constraint:single-car-4}\\
b^i_{s_{k}} &\geq b_{min}, \quad s_{k} \in\left\{j | x_{j}^{i}=1\right\},				\label{constraint:single-car-5}\\
b^i_{d} &\geq b_{m i n},																	\label{constraint:single-car-6}\\
(1-\omega) \widetilde{f}_{j} &\leq f_{j} \leq(1+\omega) \widetilde{f}_{j}.				\label{constraint:single-car-7}
\end{align}
where $\{j | x_{j}^{i}=1 \}$ means the chosen WCLs. $b_{max}$ and $b_{min}$ denote the maximum SOC of EVs and lower bound of SOC requirement respectively, which are the same for all EVs. \eqref{constraint:single-car-3} and \eqref{constraint:single-car-4} limit EV charging capacity not to exceeding WCL maximum charge quantity and its maximum SOC, respectively. \eqref{constraint:single-car-5} and \eqref{constraint:single-car-6} ensure EV can arrive each chosen WCL and destination. \eqref{constraint:single-car-7} means that the traffic flow of WCL is near the predicted value. It should be noted that although EVs do not consider constraint \eqref{constraint:single-car-7}, as the game coordinator, the traffic center will impose this constraint to the lower layer game to maintain the stability of traffic flow.

Besides, it should be noted that $b^i_j$ is the SOC of an EV at WCL $j$, which changes with different WCL sequence. The dynamic SOC model is:
\begin{equation}
\label{eq:dynamic-model}
\left[\begin{array}{c}
{b^i_{s}} \\
{b^i_{s_{1}}} \\
{\vdots} \\
{b^i_{s_{n}}} \\
{b^i_{d}}
\end{array}\right]=\left[\begin{array}{c}
{b^i_{s}} \\
{b^i_{s}-\bar{p}_{c}L\left(s, s_{1}\right)} \\
{\vdots} \\
{b^i_{s_{n-1}}-\bar{p}_{c}L\left(s_{n-1},s_{n}\right)+u_{s_{n-1}}^{i}} \\
{b^i_{s_{n}}-\bar{p}_{c}L\left(s_{n}, d\right)+u_{s_{n}}^{i}}
\end{array}\right],
\end{equation}
where $s_k$ is a WCL in an ordered charging sequence.
\subsubsection{Potential Function Construction}
For the convenience of analysis, we divide \eqref{obj:single-car} into three parts:
\begin{align}
J_{E V}^{1}\left(a_{i}, a_{-i}\right)&=\sum_{j \in \left\{j | x_{j}^{i}=1\right\}} P\left(U^F_{j}\right) u_{j}^{i},\\
J_{E V}^{2}\left(a_{i}, a_{-i}\right)&=\sum_{j \in \left\{j | x_{j}^{i}=1\right\}} D\left(f_{j}\right),\\
J_{E V}^{3}\left(a_{i}, a_{-i}\right)&=\sum_{j \in \left\{j | x_{j}^{i}=1\right\}} R\left(u_{j}^{i}\right) \nonumber\\
&+C\left(s_{i}, X_{i}, d_i\right)+T\left(s_{i}, X_{i}, d_i\right).
\end{align}
where $a_i = \{X_i,u^i_j, j \in J\}$. The interaction of charging cost and traffic congestion between EVs are reflected in $J_{E V}^{1}$ and $J_{E V}^{2}$. $J_{E V}^{3}$ is only about the loss of an EV itself. To show the lower layer game is a potential game, we define the corresponding functions:
\begin{align}
\label{eq:potential-func-1}
\Phi^{1}\left(a_{i}, a_{-i}\right)&=\sum_{j \in J} P\left(U^F_{j}\right) U^F_{j},\\
\label{eq:potential-func-2}
\Phi^{2}\left(a_{i}, a_{-i}\right)&=\sum_{j \in J} \sum_{k=1}^{f_{j}} D(k),\\
\label{eq:potential-func-3}
\Phi^{3}\left(a_{i}, a_{-i}\right)&=\sum_{i \in I} \left[C\left(s_{i}, X_{i}, d_{i}\right)+ T\left(s_{i}, X_{i}, d_{i}\right)\right] \nonumber\\
&-\sum_{i \in I}\sum_{j \in \left\{j | x_{j}^{i}=1\right\}} R\left(u_{j}^{i}\right),
\end{align}

Next we consider the change in the $J_{E V}^{i}$ and $\Phi^{i}$ due to an action change of EV $i$. We can write: 

\begin{equation}
\begin{aligned}
J_{E V}^{1}&\left(a_{i}, a_{-i}\right)-J_{E V}^{1}\left(a_{i}^{\prime}, a_{-i}\right)\\
& =\sum_{a_i \backslash a_{i}^{\prime}}\left[P(U^F_j)+q u_j^{\prime i}\right]\left(u_{j}^{i}-u_{j}^{\prime i}\right),\\
\Phi^{1}&\left(a_{i}, a_{-i}\right)-\Phi^{1}\left(a_{i}^{\prime}, a_{-i}\right)\\
&=\sum_{a_i \backslash a_{i}^{\prime}}\left[P(U^F_j) + q(U^F_j+u_{j}^{\prime i}-u_j^i) \right]\left(u_{j}^{i}-u_{j}^{\prime i}\right).\\
J_{E V}^{2}&\left(a_{i}, a_{-i}\right)-J_{E V}^{2}\left(a_{i}^{\prime}, a_{-i}\right)\\
&=\sum_{j \in X_{i} \backslash X_{i}^{\prime}} D\left(f_{j}\right)-\sum_{j \in X_{i}^{\prime} \backslash X_{i}} D\left(f_{j}+1\right),\\
\Phi^{2}&\left(a_{i}, a_{-i}\right)-\Phi^{2}\left(a_{i}^{\prime}, a_{-i}\right)\\
&=\sum_{j \in X_{i} \backslash X_{i}^{\prime}} D\left(f_{j}\right)-\sum_{j \in X_{i}^{\prime} \backslash X_{i}} D\left(f_{j}+1\right),\\
J_{E V}^{3}&\left(a_{i}, a_{-i}\right)-J_{E V}^{3}\left(a_{i}^{\prime}-a_{-i}\right)\\
&=\Phi^{3}\left(a_{i}, a_{-i}\right)-\Phi^{3}\left(a_{i}^{\prime}, a_{-i}\right)
\end{aligned}
\end{equation}
Finally noting that $\Phi = \Phi^{1} + \Phi^{2} + \Phi^{3}$ and $J_{EV} = J_{EV}^1 + J_{EV}^2 + J_{EV}^3$, if $J_{EV}(a_i,a_{-i}) - J_{EV}(a_i^{\prime},a_{-i}) > 0$, we have $\Phi(a_i,a_{-i}) - \Phi(a_i^{\prime},a_{-i}) > 0$, This shows that $\Phi(\cdot)$ is an ordinal potential function for the game, and hence, it admits a pure-strategy Nash Equilibrium (NE) \cite{MONDERER1996124Potential}.

\subsection{Upper Layer: Game Between WCLs and EVs}
As for the upper layer game, we use reverse Stackelberg game to model it. This game is a dynamic game with incomplete information, which is applicable to the situation that the leader and the follower know few about each other. The leader can obtain the follower's information by informing followers about leaders' decision function of the follower decision, so as to stimulate the follower to carry out certain cooperation. In our problem, WCLs are leaders and EVs are followers. WCLs guide EVs through the price function, so that the total charge of EVs is close to their expected value. 

\subsubsection{Reverse Stackelberg Game Process}
\begin{itemize}
	\item WCLs broadcast their price function $P(U_j)$ to EVs,
	\item According to the price function, EVs give their own charging plan and feed it back to WCLs;
	\item According to the charging plan of EVs, WCLs calculate the actual electricity price according to the previously determined price function, and inform all vehicles;
	\item All EVs are charged according to their charging plans, and the corresponding fees shall be paid.
\end{itemize}

\subsubsection{WCLs utility funciton}
As the electricity seller, WCLs have two objectives, increasing electricity sales and decreasing additional loss caused by the deviation of actual sold electricity and predicted quantity. Hence the utility function is as follows:

\begin{align}
\max_{U^L_j,q_j}&\qquad J_{W C L}=\sum_{j \in J}\left[P\left(U^L_{j}\right) U^L_{j}-\mu\left(U^L_{j}-\widetilde{U}_{j}\right)^{2}\right] \label{obj:wcl}\\
s.t.&\qquad\qquad\qquad\quad\  U^L_j \ge 0, \label{constraint:wcl-1}\\
&\qquad\qquad\qquad P(U^L_j) \ge 0, \label{constraint:wcl-2}
\end{align}
where $\mu$ is the additional cost coefficient, which is larger than price coefficient $q_j$.

It should be noted that the actual decision of the WCL $j$ is $q_j$. $U^L_j$ is the desired electricity sale of the WCL, while the actual sale $U^F_j$ is determined by EVs rather than the WCL. Therefore, the best $q_j$ that guide $U ^ F_j$ to approach $U ^ L_j$ cannot be obtained from \eqref{obj:wcl}. Through the upper layer game process, the leader continuously changes $q_j$ to guide the follower decision, such that the final $U^F_j$ is consistent with $U^L_j$. This principle helps us to design the iteration algorithm for reverse stackelberg game in \ref{subsec:reverse}.

%Although the leader WCLs hope to stimulate the EV decision-making by the price coefficient $q_j$ to make $U^F_j$ reach the desired electricity sales, the direct relationship between $q_j$ and $U^F_j$ cannot be obtained in leader optimization. Therefore, for \eqref{obj:wcl}, we regard $U^L_j$ as variables under given $q_j$ to find the best desired electricity sales of WCLs. Through the upper layer game process, we continuously optimize $q_j$ so that the best $U^F_j$ of follower EVs and the desired $U^L_j$ of leader WCLs can be consistent. This principle helps us to design the iteration algorithm for reverse stackelberg game later.

%However, WCLs can only stimulate EV decision-making by changing the price coefficient $q_j$, and can not directly control the electricity sales. Therefore, in the end of the upper layer game, the charging quantity in leader optimization and the one in follower optimization should be the same. This principle helps us to design the iteration algorithm for reverse stackelberg game later.
\section{Algorithm}
\label{sct4}
\subsection{TGSP Based Shortest Charging Sequence Algorithm}
\label{subsec:tgsp}
We have mentioned shortest charging route length function $L()$ in \ref{sec:price-function}. In this subsection, we will introduce this function, which is obtained by TGSP based shortest charging sequence algorithm. The algorithm process is as follows:
\begin{enumerate}[i]
	\item \textit{Build Distance Matrix:} Since we know the node connections of the road network and the distance between adjacent nodes, we choose \textit{Floyd} algorithm to calculate the node distance matrix, which contains the distance between any two intersections. 
	\item \textit{Determine charging sequence:} To find the length of shortest path going through chosen WCLs, we have to determine the sequence of WCLs. With $n$ chosen WCLs, the sequence permutation complexity is $O(n!)$. As $n$ increases, it will cost much time to find the best sequence. Because the vehicle charging route is generally towards the end point, based on the greedy idea, we search the shortest sequence in three ways: forward searching, backward searching and searching from both ends to the middle. The algorithm is shown in Algorithm \ref{a:tgsp}.
	\item \textit{Shortest Length Calculation:} From the results of previous step, the shortest charging path length can be obtained as $min\{Len1,Len2,Len3\}$.
\end{enumerate}
\begin{algorithm}[h]
	\caption{TGSP}
	\label{a:tgsp}
	\begin{algorithmic}[1]
		\REQUIRE Distance matrix $G^*$, the start and destination of EVs $s,d$, chosen WCLs $X$, the beginning and end intersection of WCLs $b,e$. 
		\ENSURE Shortest charging sequence $S$ and length $Len$.
		\STATE Initialize sequence $S1,S2,S3$, length $Len1,Len2,Len3$.
		\STATE \textbf{Forward searching:} $last = s$
		\WHILE {Some chosen WCLs are not sorted in S1 }
		\FOR {Each $j \in$ residual WCLs}
		\STATE $dist = G^*(last,b_j)+G^*(b_j,e_j), last = e_j$,
		\ENDFOR
		\STATE put the WCL with the shortest $dist$ into $S1$ in order, 
		\STATE$Len1 = Len1 + dist$.
		\ENDWHILE
		\STATE \textbf{Backward searching:} $last = d$
		\WHILE {Some chosen WCLs are not sorted in S2 }
		\FOR {Each $j \in$ residual WCLs}
		\STATE $dist = G^*(last,e_j)+G^*(b_j,e_j), last = s_j$,
		\ENDFOR
		\STATE put the WCL with the shortest $dist$ into $S2$ reversely,
		\STATE $Len2 = Len2 + dist$.
		\ENDWHILE
		\STATE \textbf{Searching from both ends to the middle:} $last_l = s,last_r = d$
		\STATE $num=\sum X$, $num_{half} = num/2$ rounded down. 
		\FOR {$k=num_{half}$}
		\FOR {$j \in$ residual WCLs}
		\STATE $dist = G^*(last,b_j)+G^*(b_j,e_j), last_l = e_j$,
		\ENDFOR
		\STATE put the WCL with the shortest $dist$ into $S3$ in order, 
		\STATE$Len3 = Len3 + dist$.
		\FOR {Each $j \in$ residual WCLs}
		\STATE $dist = G^*(last,e_j)+G^*(b_j,e_j), last_r = s_j$,
		\ENDFOR
		\STATE put the WCL with the shortest $dist$ into $S3$ reversely,
		\STATE $Len3 = Len3 + dist$.
		\ENDFOR
		\IF {$num\%2==0$}
		\STATE $Len3 = Len3 + G^*(last_l,last_r)$,
		\ELSE 
		\STATE the last WCL $j$,$Len3 = Len3 + G^*(last_l,b_j) + G^*(b_j,e_j) + G^*(e_j,last_r)$
		\ENDIF
		\RETURN $Len = min\{Len1,Len2,Len3\}$ and corresponding sequence $S$.
	\end{algorithmic}
\end{algorithm}
\subsection{Improved PSO}
%Combining \eqref{eq:potential-func-1},\eqref{eq:potential-func-2} and \eqref{eq:potential-func-3}, we get the potential function and constrains as:
%\begin{align}
%\label{obj:follower}
%\min&\quad \sum_{j \in J} P\left(U_{j}\right) U_{j}+\sum_{i \in I}\left[C\left(s_{i}, X_{i}, d_{i}\right)+T\left(s_{i}, X_{i}, d_{i}\right)\right]\nonumber\\
%&\quad-\sum_{i \in I}\sum_{j \in \left\{j | x_{j}^{i}=1\right\}} R\left(u_{j}^{i}\right)+\sum_{j \in J} \sum_{k=1}^{f_{j}} D(k)\\
%s.t.&\qquad\qquad\qquad\qquad\qquad\qquad \eqref{constraint:single-car-1} \sim \eqref{constraint:single-car-7}. \nonumber
%\end{align}

In the lower layer potential game, the change trend of potential function value is the same as that of the individual cost. Finding the optimal decision for the potential function is equivalent to finding the best decision for each EV. In this problem, the traffic center, as the game coordinator, is responsible for calculating and notifying the results of vehicle charging planning. 
As potential function $\Phi$ makes the EV-EV game a Mixed Integer NonLinear Programming (MINLP) problem and the dynamic SOC of EVs needs to be calculated after the charging sequence is determined, we design an improved PSO algorithm to solve this problem. First, we divide $\Phi$ into two parts: the nonlinear programming of charging quantity planing with fixed $X_i$ $P^1$, and the integer programming of WCL selection $P^2$:
\begin{align}
\label{obj:follower-nlp}
P^1:\min_{U^i}& \quad obj^1(U^i|X_i^*)=\sum_{j \in J} P\left(U^F_{j}\right) U^F_{j}\nonumber\\
&\qquad\qquad\qquad\quad\   - \sum_{i \in I}\sum_{j \in \left\{j | x_{j}^{i}=1\right\}} R\left(u_{j}^{i}\right)\\
s.t.&\qquad\qquad \eqref{constraint:single-car-1},\eqref{constraint:single-car-3} \sim \eqref{constraint:single-car-6}. \nonumber
\end{align}
\begin{align}
\label{obj:follower-ip}
P^2:\min_{X_i}& \quad obj^2(X_i)=\sum_{i \in I}\left[C\left(s_{i}, X_{i}, d_{i}\right)+T\left(s_{i}, X_{i}, d_{i}\right)\right]\nonumber\\
&\qquad\qquad\quad\quad\  +\sum_{j \in J} \sum_{k=1}^{f_{j}} D(k)\\
s.t.&\qquad\qquad\qquad\qquad \eqref{constraint:single-car-2},\eqref{constraint:single-car-7}. \nonumber
\end{align}

By calculating the Hessian matrix, it is easy to show that $P^1$ is a convex problem. Due to the dynamic SOC constrains \eqref{constraint:single-car-4} $\sim$ \eqref{constraint:single-car-6}, it's hard to get the general form of solution, but can be solved by solvers like CPLEX. The traditional PSO algorithm is not good at solving integer programming\cite{Wu2019Integer}. Hence, to solve $P^2$, we make following improvements:
\begin{itemize}
	\item \textit{Position Updating:} When the iteration is more than a certain number of times, the particles are nearly stable. In order to accelerate convergence, we update particles' velocity and position in the following way:
	\begin{equation}
	\label{eq:PSO-velocity-2}
	V_{i}=C_{1} rand(0,1)\left(P_{i}-X_{i}\right)+C_{2} rand(0,1)\left(P_{g}-X_{i}\right),
	\end{equation}
	\begin{equation}
	\label{eq:sigmoid-2}
	s=Sigmoid(V_i)=\left\{\begin{array}{ll}
	{1-\frac{2}{1+\exp \left(-V_i\right)}}, & {\text { if } V_i \leq 0}, \\
	{\frac{2}{1+\exp \left(-V_i\right)}-1}, & {\text { if } V_i>0}.
	\end{array}\right.
	\end{equation}
	\begin{equation}
	\label{eq:PSO-position-2-1}
	if\quad V_i<0,\qquad
	X_{i}=\left\{\begin{array}{ll}
	{0} & {\text { random(0,1) } < s,} \\
	{X_i} & {others.}
	\end{array}\right.
	\end{equation}
	\begin{equation}
	\label{eq:PSO-position-2-2}
	if\quad V_i>0,\qquad
	X_{i}=\left\{\begin{array}{ll}
	{1} & {\text { random(0,1) } < s,} \\
	{X_i} & {others.}
	\end{array}\right.
	\end{equation}
	$X_i$ and $V_i$ are the position and velocity of particles respectively. $P_i$ and $P_g$ are the best individual position and group position respectively. \eqref{eq:PSO-velocity-2} means that at the later stage of iteration, particles only approach to the best position of individuals and populations. \eqref{eq:sigmoid-2} $\sim$ \eqref{eq:PSO-position-2-2} reduces the possibility of large changes in the position of low-speed particle, and improves the probability of approaching the historical best position for high-speed particles.
	\item \textit{Constrains Handling:} When particles search, they may exceed the constraints \eqref{constraint:single-car-7}. We add the overstep value as a penalty term to particles' fitness. The penalty term is as follows:
	\begin{equation}
	\label{eq:overstep}
	\begin{split}
		\theta(X_i) = &\max\{0,\sum_{i \in I}x^i_j-(1+\omega)\widetilde{f}_{j}\}\\
		 &+ \max\{0,(1-\omega)\widetilde{f}_{j}-\sum_{i \in I}x^i_j\},
	\end{split}
	\end{equation}
	\begin{equation}
	\label{eq:fitness}
	fit(X_i) = obj^1(U^{i*}|X_i)+obj^2(X_i)+\lambda \theta(X_i),
	\end{equation}
	where $\lambda$ is a large constant.
\end{itemize}
\begin{figure*}[t]
	\begin{minipage}[t]{0.33\linewidth}
		\centerline{\includegraphics[width=1\textwidth]{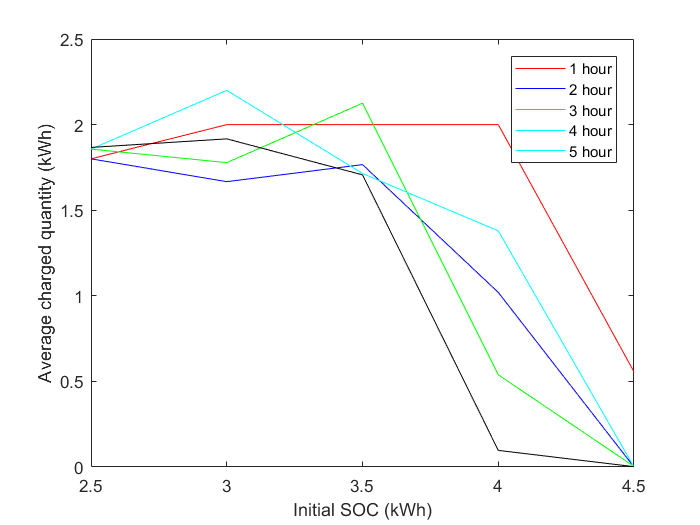}}
		\setlength{\abovecaptionskip}{-1pt}
		\caption{Average charged quantity.}
		\label{fig:ev_charge_hourly}
	\end{minipage}
	\begin{minipage}[t]{0.33\linewidth}
		\centerline{\includegraphics[width=1\textwidth]{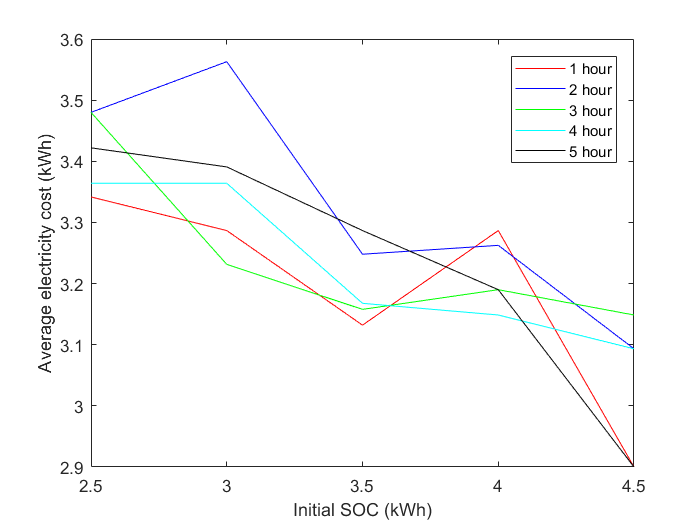}}
		\setlength{\abovecaptionskip}{-1pt}
		\caption{Average electricity cost.}
		\label{fig:ev_cost_hourly}
	\end{minipage}
	\begin{minipage}[t]{0.33\linewidth}
		\centerline{\includegraphics[width=1\textwidth]{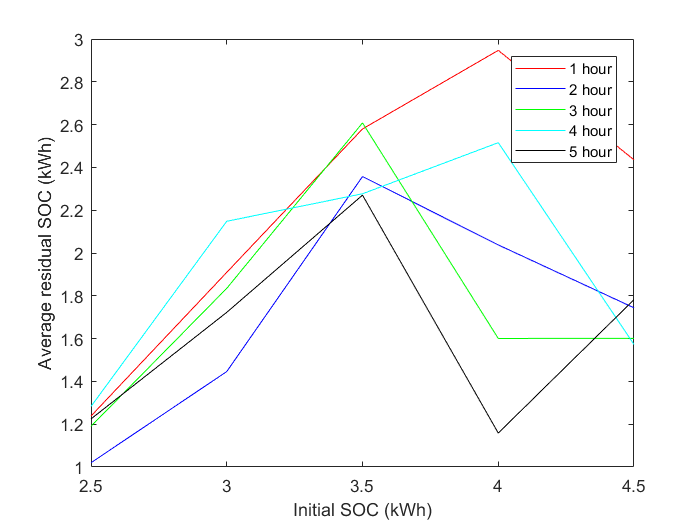}}
		\setlength{\abovecaptionskip}{-1pt}
		\caption{Average residual SOC.}
		\label{fig:ev_residual_hourly}
	\end{minipage}

	\begin{minipage}[t]{0.33\linewidth}
		\centerline{\includegraphics[width=1\textwidth]{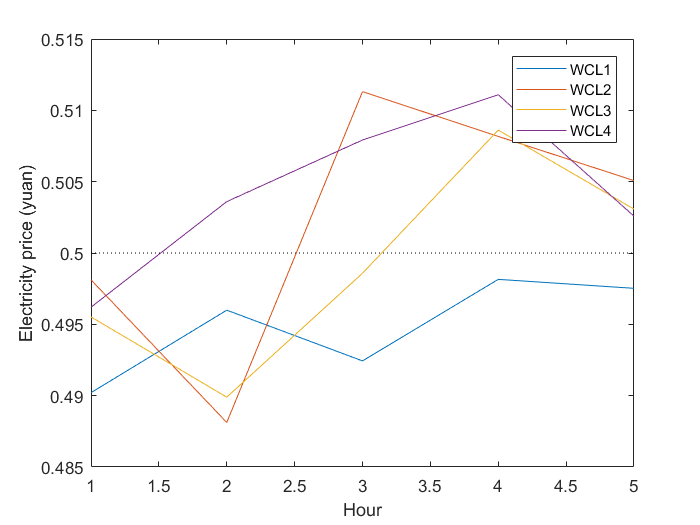}}
		\setlength{\abovecaptionskip}{-1pt}
		\setlength{\belowcaptionskip}{-1cm}   
		\caption{The charging price.}
		\label{fig:wcl_price}
	\end{minipage}
	\begin{minipage}[t]{0.33\linewidth}
		\centerline{\includegraphics[width=1\textwidth]{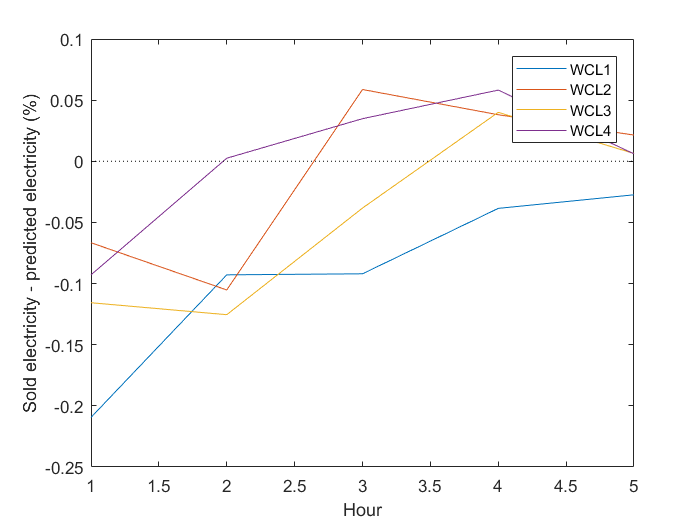}}
		\setlength{\abovecaptionskip}{-1pt}
		\setlength{\belowcaptionskip}{-1cm}  
		\caption{The total sold electricity.}
		\label{fig:wcl_charge}
	\end{minipage}
	\begin{minipage}[t]{0.33\linewidth}
		\centerline{\includegraphics[width=1\textwidth]{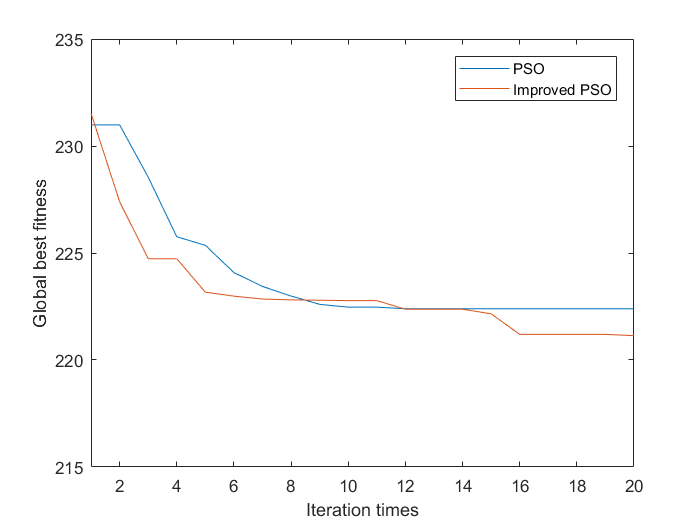}}
		\setlength{\abovecaptionskip}{-1pt}
		\setlength{\belowcaptionskip}{-1cm}  
		\caption{PSO convergence.}
		\label{fig:pso_convergence}
	\end{minipage}
\end{figure*}
\subsection{Reverse Stackelberg Game Iteration}
\label{subsec:reverse}
When the game reach equilibrium point, the total charge quantity in follower and leader optimization should be the same. Based on this, we design the iterative algorithm as follows:
\begin{enumerate}[i]
	\item Initialize electricity price coefficient $Q_0 = \{q_{j,0}|j\in J\}$;
	\item Bring $Q_{j,k-1}$ (or $Q_{j,0}$) into lower layer potential game, and get the best strategy $\{X^*_{i,k}|i\in I\}$ and $\{U^{F*}_{j,k}|j\in J\}$;
	\item Bring $Q_{j,k-1}$ (or $Q_{j,0}$) into \eqref{obj:wcl}, and get the best strategy $\{U^{L*}_{j,k}|j\in J\}$;
	\item Update the price coefficient $q_{j,k}$ with the difference of $U^{F*}_{j,k}$ and $U^{L*}_{j,k}$ as follows, to reduce the difference\cite{WEI2017315Stackelberg}:
	\begin{align}
	q_{j, k+1}&=q_{j, k}+l\left(U_{j,k}^{F*}-U_{j,k}^{L*}\right),\label{eq:iter-q}\\
	l&=\dfrac{1}{\varphi+k},
	\end{align}
	where $l$ is the learning rate that decreases with the number of iterations $k$, and $\varphi$ is a positive constant.
	\item Check the convergence condition:
	\begin{equation}
	\label{eq:iter-converge}
	|U^{F*}_{j,k} - U^{L*}_{j,k}| \le \sigma.
	\end{equation}
	If \eqref{eq:iter-converge} is satisfied, output the final follower schedule $\{X^*_{i,k}|i\in I\}$ and $\{U^{F*}_{j,k}|j\in J\}$, and leader strategy $Q_k$; otherwise continue iteration $k = k+1$ until reach the iteration limit.  
	
\end{enumerate}
\section{Performance Evaluation}
\label{sct5}
\subsection{Experiment Settings}
In this paper, SUMO is used for simulation. For the convenience of analysis, we build a road network which includes 50 roads of 1km in length. 4 roads are selected to lay the WCLs of 40m in length. $b_{max} = 5$kWh, $b_{min}=0.5$kWh, initialized SOC $b^i_0$ is randomly assigned between 2.5kWh $\sim$ 5kWh. $\omega=0.2$, $p_0=0.5$, $\alpha=0.01$, $\sigma=1$, $\varphi=500$. EVs departure from the southwest corner to the northeast corner within 5 hours, and the traffic flow gradually increases from 24 per hour to 60 per hour.
\subsection{Experiment Results}
\subsubsection{Algorithm Effect}
\begin{itemize}
	\item \textit{TGSP:} We choose 4 $\sim$ 8 roads to install WCLs in the road network, and observe the length of charging sequence route calculated by TGSP compared with the shortest route after traversing all sequences. From TABLE \ref{tb:tgsp}, we can see that TGSP can obtain basically the same short charging sequence in a shorter time.
	\begin{table}[h]
		\setlength{\abovecaptionskip}{0cm} 
		\caption{Comparison of TGSP and Traversal}
		\begin{center}
			\begin{tabular}{p{0.8cm}<{\centering} p{1cm}<{\centering} p{2cm}<{\centering} p{1.2cm}<{\centering} p{1cm}<{\centering}} 
				\hline
				\textbf{WCLs} &\textbf{Methods} & \textbf{Sequence}& \textbf{Length}(km) & \textbf{Time}(ms)\\
				\hline
				\multirow{2}{*}{4}& \textit{Traversal}& [1,3,2,4]& 18 & 0.2\\
				&\textit{TGSP}& [1,3,4,2]& 18& 0.2\\
				\hline
				\multirow{2}{*}{5}& \textit{Traversal}& [1,3,5,2,4]& 18 & 0.4\\
				&\textit{TGSP}& [1,3,5,4,2]& 18 & 0.2\\
				\hline
				\multirow{2}{*}{6}& \textit{Traversal}& [1,3,5,4,6,2]& 20 & 1.1\\
				&\textit{TGSP}& [1,3,5,4,6,2]& 20 & 0.3\\
				\hline
				\multirow{2}{*}{7}& \textit{Traversal}& [1,3,5,6,2,4,7]& 20 & 3.5\\
				&\textit{TGSP}& [1,3,5,6,2,4,7]& 20 & 0.4\\
				\hline
				\multirow{2}{*}{8}& \textit{Traversal}& [6,1,3,5,7,2,4,8]& 20 & 25.2\\
				&\textit{TGSP}& [3,5,7,2,6,1,4,8]& 22 & 0.5\\
				\hline
			\end{tabular}
			\label{tb:tgsp}
		\end{center}
	\end{table}
	\item \textit{Improved PSO:} In order to reduce the randomness of the experiment, we have carried out the same PSO process five times, and take the average value. Fig.~\ref{fig:pso_convergence} shows that, the improved PSO converges faster than the traditional PSO, and it is more likely to get lower fitness because particles search near the historical best value rather than moving randomly and substantially in the later stage.
\end{itemize}
\subsubsection{EVs Results}
\begin{itemize}
	\item \textit{Average charge quantity:} From Fig.~\ref{fig:ev_charge_hourly} we can see that, with the increase of initial SOC, the average charge of EVs decreases. This is because EVs with lower initial power has higher charging demand to ensure the endurance. EVs with sufficient electricity will choose road without the WCL to avoid traffic congestion. 
	\item \textit{Average energy cost:} Fig.~\ref{fig:ev_cost_hourly} shows that EVs with higher initial SOC cost fewer energy, which indicates that they can choose the shortest path to reach the destination, while EVs with lower initial SOC may take a detour to find a WCL to maintain their SOC.
	\item \textit{Average residual energy:} Fig.~\ref{fig:ev_residual_hourly} shows that EVs with different initial SOC are able to maintain sufficient power after finishing the trip.
\end{itemize}
\subsubsection{WCLs Results}
\begin{itemize}
	\item \textit{Electricity price:} Fig.~\ref{fig:wcl_price} shows that when the traffic flow is low, the WCL attract EVs to charge with low price. As the traffic flow increases, the electricity price increases to discourage some EVs with weak charging demand.
	\item \textit{Deviation of charge quantity:} Fig.~\ref{fig:wcl_charge} indicates that when the traffic flow increases, it is easier for WCLs to guide EVs' total charge approach the predicted value.
\end{itemize}
\section{Conclusion}
\label{sct6}
In this paper, to find a good charging schedule for WCLs and EVs, we describe the relationship between power grid, WCLs and EVs to illustrate the mutual benefit impact. Then we define the important models, including the price function of WCLs, the maximum charging quantity model and loss models of EVs. In order to coordinate the income balance between EVs, we use the potential game model to prove the existence of NE in the EV-EV charging game, and design the TGSP algorithm to solve the dynamic charging sequence, as well as the improved PSO algorithm to solve the MINLP potential game. In order to balance the benefits of WCLs and EVs, we use the reverse Stackelberg game model to help WCLs stimulate EVs' charging plan through the price function, and design the iteration algorithm to get the optimal decision of both sides. The numerical simulation results show that the charging sequence obtained by TGSP algorithm is consistent with the shortest sequence, and the improved PSO algorithm has better convergence. The proposed double-layer game model can achieve a good balance effect on the benefits of both WCLs and EVs. 
Through a reasonable price adjustment scheme, WCLs can achieve a higher balance of electricity sales, and EVs with different SOC can complete the process of supplying electricity under the condition of lower energy consumption.

\section*{Acknowledgment}
This work was supported by National Natural Science Foundation of China (61731012, 61573245 and 61933009).
\balance
\bibliographystyle{IEEEtran}
\bibliography{refs}
%\begin{thebibliography}{00}
%	\bibitem{b8}
%	I.~Hwang, Y.~J. Jang, Y.~D. Ko, and M.~S. Lee, ``System optimization for
%	dynamic wireless charging electric vehicles operating in a multiple-route
%	environment,'' \emph{IEEE Transactions on Intelligent Transportation
%		Systems}, vol.~19, no.~6, pp. 1709--1726, 2018.
%\end{thebibliography}
\vspace{12pt}
\color{red}
\end{document}